%% file: main.tex
\documentclass[conference]{IEEEtran}

\usepackage{algpseudocode,algorithm}
\usepackage{multirow}
\AtBeginDocument{%
  }
\usepackage{graphicx}
\usepackage{subcaption}
\usepackage{lipsum}
\usepackage{physics}
\usepackage{quantikz}
\usepackage{amsfonts} 
\usepackage{adjustbox}

\usepackage{balance}

\begin{document}

\title{\huge quAssert: Automatic Generation of Quantum Assertions}

 \author{Hasini Witharana, Daniel Volya and Prabhat Mishra \\ 
 University of Florida, Gainesville, Florida, USA}



\maketitle

\begin{abstract}
Functional validation is necessary to detect any errors during quantum computation. There are promising avenues to debug quantum circuits using runtime assertions. However, the existing approaches rely on the expertise of the verification engineers to manually design and insert the assertions in suitable locations. In this paper, we propose automated generation and placement of quantum assertions based on static analysis and random sampling of quantum circuits. Specifically, this paper makes two important contributions. We automatically uncover special properties of a quantum circuit, such as purely classical states, superposition states, and entangled states using statistical methods. We also perform automated placement of quantum assertions to maximize the functional coverage as well as minimize the hardware overhead. We demonstrate the effectiveness of the generated assertions in error detection using a suite of quantum benchmarks, including Shor's factoring algorithm and Grover's search algorithm.

\end{abstract}

\pagestyle{empty}

\input{section/intro.tex}
\input{section/related.tex}
\input{section/assertion.tex}
\input{section/generation.tex}

\input{section/experiment.tex}
\input{section/conclusion.tex}

\section{Acknowledgments}

This work was partially supported by the grant from Semiconductor Research Corporation (2020-CT-2934).

\balance

\bibliographystyle{IEEEtran}
\bibliography{reference.bib}

\end{document}

%% file: section/intro.tex
\section{Introduction}
\label{sec:introduction}

Assertions provide a mechanism to describe desirable properties of a system that should be satisfied. Assertion-based validation is widely used for both pre-silicon and post-silicon validation in classical systems~\cite{witharanaSurveyAssertionbasedHardware2022}. Similar to classical computation, bugs or errors \cite{volyaSpecialSessionImpact2020} can be present in a quantum computation, constituting a need for verifying quantum programs \cite{huangQDBQuantumAlgorithms2019, amyLargescaleFunctionalVerification2019}. \textit{Quantum assertions} is a promising avenue for validating a quantum program's functionality \cite{garciadelabarreraQuantumSoftwareTesting, wederQuantumSoftwareDevelopment}.  We specifically use the term \textit{quantum assertions} to denote assertions that assert properties of a \textit{quantum state}, and therefore should  be evaluated as part of quantum computation at run time. To enable assertion-based validation of quantum systems, we need to answer two important questions: \textit{how to generate and where to place the quantum assertions?}  

\begin{figure}[t]
	\centering
\includegraphics[width=1\columnwidth]{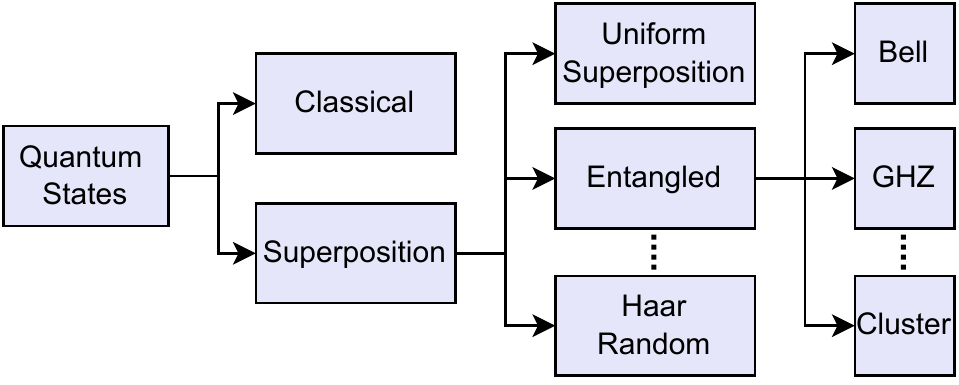}
 \vspace{-0.2in}
	\caption{A quantum state can be either in a single computational state or a general combination of computational states (classical and superposition). The superposition state can be further divided in various categories as outlined in Section~\ref{sec:states}.}
	\label{fig:classification}
	    \vspace{-0.2in}
\end{figure}

Recent approaches investigate implementations for quantum assertions~\cite{zhouQuantumCircuitsDynamic,huangStatisticalAssertionsValidating2019, liProqProjectionbasedRuntime2020}. 
There are few practical limitations of the existing methods. Quantum assertion generation is an entirely manual process that requires expert knowledge to craft effective assertions that utilize minimal quantum resources and provides reasonable functional coverage. It also relies on the expertise of the designers to identify the placement of these assertions. Moreover, quantum computations are ``parallel'' in nature due to superposition. A general quantum state may consist of many computational states, as shown in Figure~\ref{fig:classification}. Section~\ref{sec:states} provides a comprehensive survey of these quantum states. A quantum operation will act on all these computational states. Furthermore, superposition enables \textit{entanglement} whereby measurement of one subsystem reveals the state of the other subsystem. The complexity of quantum states coupled with systems with large number of qubits makes it infeasible to manually generate effective assertions.

In this paper, we propose an automated framework for generation of quantum assertions that consists of the following tasks. First, the quantum circuit is statically analyzed to uncover common functionalities. Next, the qubits and gates are randomly sampled from a quantum circuit with respect to the discovered functionalities. Then, the properties of the gates are sorted into three categories of assertions: classical, uniform superposition, or entanglement using classical simulation of selected sub circuit of interest and statistical analysis. Finally, the placement of assertions is done based on the  assertion type and the results of random sampling of the circuit. Specifically, this paper makes the following contributions:
\begin{itemize}
    \item Surveys quantum states of interest to define three classes of quantum assertions.
    \item Proposes an automated algorithm for generation of quantum assertions using static analysis and random sampling.
    \item Demonstrates the utility of these assertions in verifying quantum circuits.
\end{itemize}

This paper is organized as follows. Section~\ref{sec:background} provides relevant background and surveys related efforts. Section~\ref{sec:classes} defines three major classes of quantum assertions. Section~\ref{sec:assertion} describes our automated assertion generation framework. Section~\ref{sec:experiments} presents experimental results. Finally, Section~\ref{sec:conclusion} concludes the paper.




    


%% file: section/related.tex
\section{Background and Related Work}
\label{sec:background}
This section first surveys different quantum states. Next, it presents the related work on quantum assertions.

\subsection{Survey of Quantum States}
\label{sec:states}

As shown in Figure~\ref{fig:classification}, a quantum state can be in two classes of states: classical and superposition~\cite{nielsen2000quantum}. The superposition state can be further classified into multiple categories including uniform, entangled, random, etc. The remainder of this section briefly describes different quantum states.

\vspace{0.05in}
\noindent \textbf{Classical States:} A quantum state that is in only one of the basis states is considered a classical state. For $n$-qubits, a classical state has the form, up to a global phase $\theta$,
\vspace{-0.05in}
$$\ket{\psi_C} = e^{i\theta} \sum_{i=0}^{n-1} \delta_{i,k} \ket{i} = e^{i\theta}\ket{k}
\vspace{-0.05in}
$$
where $\delta_{i,k}$ is the Kronecker delta denoting that only the $k$-th state is active.
Classical states often occur in encoding classical information to a quantum computer, or in classical operations such as addition or comparison.  

\vspace{0.05in}
\noindent \textbf{Uniform Superposition:} In contrast to classical states, uniform superposition states equally spread across all possible basis states. Specifically, for $n$-qubits, a uniform superposition state has the form, up to a global phase $\theta$,
\vspace{-0.1in}
$$\ket{\psi_{S}} =  \frac{e^{i\theta}}{\sqrt{2^n}}\sum_{i=0}^{n-1} \ket{i}
\vspace{-0.1in}
$$

\vspace{0.05in}
\noindent \textbf{Cat Entangled State:} If a quantum state cannot be written as a combination of individual qubit states, it is known as an entangled state. A special case of a highly-entangled state is when, for $n$-qubits, the state is a combination of all-zeros and all-ones:
\vspace{-0.05in}
$$\ket{\psi_E} = \frac{e^{i\theta}}{\sqrt{2}}\left(\ket{0}^{\otimes n} + \ket{1}^{\otimes n}\right).
\vspace{-0.05in}
$$
For two qubits and three qubits, these states are referred to as Bell and Greenberger–Horne–Zeilinger (GHZ) states, respectively.

\vspace{0.05in}





\subsection{Related Work}
There are many promising assertion generation approaches in classical domain~\cite{vasudevanGoldMineAutomaticAssertion2010}. However, they are not applicable for generating quantum assertions since output is deterministic for a given input for classical computing, while output values are a result of destructive measurements and come with a probability distribution in quantum computing. There are early efforts to discuss the importance and applicability of quantum assertions~\cite{huangStatisticalAssertionsValidating2019,liProqProjectionbasedRuntime2020}. Recent approaches explored different assertion generation methods such as ancilla-based methods \cite{zhouQuantumCircuitsDynamic}, statistical-based methods \cite{huangStatisticalAssertionsValidating2019}, and projection-based methods~\cite{liProqProjectionbasedRuntime2020}.

There are closely related efforts in quantum error correction and formal verification of quantum circuits. Error correcting codes assume a certain noise model, and provide special state encodings that can correct a state if an error is detected. Quantum assertions, although similar, are not concerned with correcting a state and only seek to assert a given property of the state. 
There are also recent efforts to check the correctness in the output of a quantum circuit, such as through formal verification of quantum circuits~\cite{randQWIREPracticeFormal2018a, burgholzer2021random}, or by assuming domain-specific knowledge (e.g., post-selection rules) to ignore incorrect outputs of a quantum computation \cite{volyaQuantumSpectralClustering2021, nguyenDigitalQuantumSimulation2022}. To the best of our knowledge, our proposed approach is the first attempt in automated generation of quantum assertions.

%% file: section/assertion.tex
\section{Classes of Quantum Assertions}
\label{sec:classes}

Any quantum algorithm will consist of a series of quantum gates that will evolve an initial quantum state to an arbitrary desired output state. Although this strategy is similar to classical computation, quantum gates must be reversible, linear, and unitary. Additionally, at any given moment in the computation, the quantum state is a general superposition (combination) of all possible states. Namely, given n-qubits initialized to $\ket{0}^{\otimes n}$, and a quantum circuit $\mathcal{U}$ consisting of m-gates in a sequence $\{U_i\}_{i=1}^m$, the output of $\mathcal{U}$ gives a final state 
$$\ket{\psi} = \mathcal{U}\ket{0}^{\otimes n} = (U_m \circ U_{m-1} \circ \hdots \circ U_1)\ket{0}^{\otimes n}.$$
Our goal is to add an assertion, $\mathcal{A}$, or even a set of assertions $\{A_i\}$, to the existing quantum circuit $\mathcal{U}$ that asserts a property of the quantum state at the given point in the quantum circuit, $\ket{\psi_i}$. For example, to assert the quantum state after the application of the first quantum gate would yield:
$$\ket{\psi^\prime} = (U_m \circ U_{m-1} \circ \hdots \circ \mathcal{A}  \circ U_1)\ket{0}^{\otimes n^\prime}.$$

\begin{figure}[htp]
	\centering
\vspace{-0.15in}
\begin{adjustbox}{width=0.6\linewidth}
	\begin{quantikz}
	    \lstick{$\ket{0}$} & \gate{H} & \qw & \gate{\mathcal{A}_1} & \ctrl{1} & \gate[wires=2]{\mathcal{A}_2} & \qw \\
	    \lstick{$\ket{0}$} & \qw & \gate[wires=3]{\mathcal{A}_0} & \qw &\targ{} & & \qw \\
	    \lstick{$\ket{0}$} & \gate{X} & \qw & \qw & \qw & \qw & \qw \\
	    \lstick{$\ket{0}$} & \gate{X} & \qw & \qw & \qw & \qw & \qw \\
	    \lstick{assert0} & \cw & \cwbend{-1} \\
	    \lstick{assert1} & \cw & \cw & \cwbend{-5} \\
	    \lstick{assert2} & \cw & \cw & \cw & \cw & \cwbend{-5}
	\end{quantikz}
\end{adjustbox}
	\caption{Three important classes of quantum assertions: classical (assert0), superposition (assert1), and entanglement (assert2).}
	\label{fig:classes}
	    \vspace{-0.15in}
\end{figure}
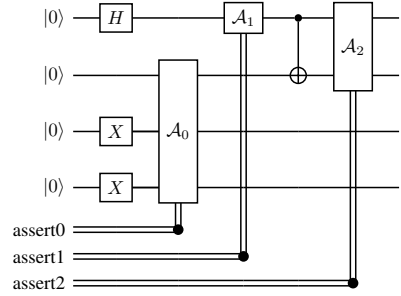

In general, there are infinitely many properties that a quantum state $\ket{\psi_i}$ may have since a quantum state is a general superposition consisting of complex-valued amplitudes. Fortunately, certain quantum states, and specific sequences of quantum gates, yield desirable and well-known states (classical, superposition, and entanglement) which are described in Section~\ref{sec:states}. In this section, we present three types of quantum assertions as shown in Figure~\ref{fig:classes} that are capable of checking these states. 

\vspace{0.05in}
\noindent \textbf{Classical Quantum Assertions:}
This assertion type has the capability of checking whether the quantum state is classical at a specific breakpoint for a given quantum design. Classical assertions can be used to debug a quantum circuit to identify whether any error/bug occurred which has the power to change the classical state to another state or change the expected classical value to some other classical value. Classical quantum assertion is composed of a combinational assertion using the propositional logic equivalent ($\Leftrightarrow$). The classical assertion template is $A : assert(\ket{input} \Leftrightarrow \ket{output})$. Use of equivalent operator checks other qualities of quantum circuit such as  reversibility. An example placement of classical assertion is shown in Figure~\ref{fig:classes} as $assert0$. 

\vspace{0.05in}
\noindent \textbf{Uniform Superposition Quantum Assertions:}
This assertion type can check whether the quantum state is uniform superposition at a specific break-point. The rest of the paper uses the term superposition assertion to represent this assertion type. Superposition assertions can be used when debugging a quantum circuit to identify any error which can change the uniform superposition state to another state. Superposition quantum assertion is composed of a combinational assertion using the propositional logic implication ($\rightarrow$). The superposition assertion template is $A : assert(\ket{input} \rightarrow \ket{\psi_{S}})$. A sample superposition assertion is shown in Figure~\ref{fig:classes} as $assert1$.

\vspace{0.05in}
\noindent \textbf{Entanglement Quantum Assertions:}
This assertion type has the capability of checking whether the quantum state is in a Cat state at a specific break-point. Entanglement assertions can be used when debugging a quantum circuit to identify whether any error occurred to change the Cat state to any other state. Entanglement quantum assertion is composed of superposition assertions using the propositional logic implication. The entanglement assertion template is $A : assert(\ket{input} \rightarrow \ket{\psi_{E}})$. A sample entanglement assertion is shown in Figure~\ref{fig:classes} as $assert2$.

%% file: section/generation.tex
\section{Quantum Assertion Generation}
\label{sec:assertion}

\begin{figure}[htp]
	\centering
\vspace{-0.1in}
\includegraphics[width=1\columnwidth]{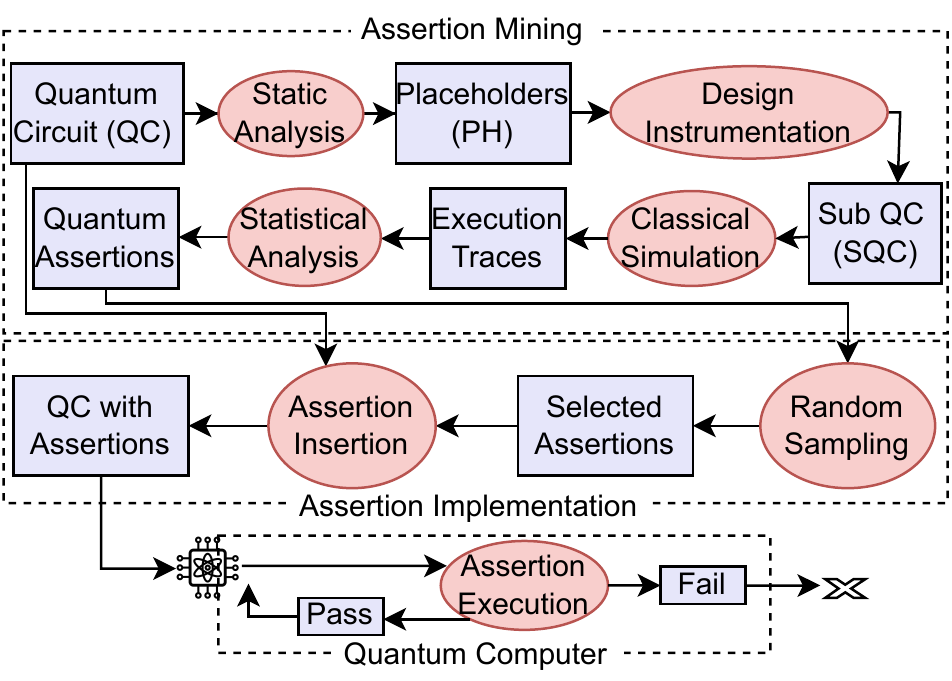}
  \vspace{-0.2in}
	\caption{An overview of our assertion generation framework that consists of two phases: assertion mining and assertion implementation.}
	\label{fig:overview}
	    \vspace{-0.1in}
\end{figure}

Figure~\ref{fig:overview} shows an overview of our proposed assertion generation framework that consists of two major phases: assertion mining and assertion implementation. We implement assertion mining in four steps. The first step statically analyzes the design and identifies the placements to measure. The second step instruments the placeholders by adding measure points to the identified spots. The third step simulates the place holders using random inputs to get the execution traces. The fourth step analyzes the traces and mines three different classes of assertions (described in Section~\ref{sec:classes}) based on the statistical analysis of the output distributions.

After a successful assertion mining process, we can proceed to assertion implementation that consists of two important steps. The first step selects assertions based on the coverage goal. The next step embeds the measurement points with a suitable assertion insertion process to the design. Finally, the quantum circuit with assertions can be executed on a real quantum machine to detect errors during run-time.
In this section, we describe both assertion mining and assertion implementation in detail.







\subsection{Assertion Mining}

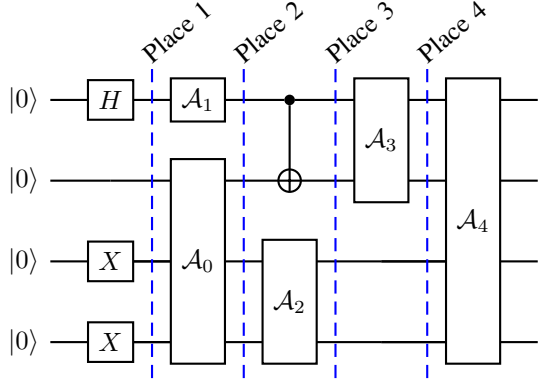
\begin{figure}[tp]
	\centering
	\begin{quantikz}[slice all,remove end slices=1,slice titles= Place \col,slice style=blue,slice label style={inner sep=1pt,anchor=south west,rotate=40}]
	    \lstick{$\ket{0}$} & \gate{H} & \gate{\mathcal{A}_1} & \ctrl{1} & \gate[wires=2]{\mathcal{A}_3} & \gate[wires=4]{\mathcal{A}_4} & \qw \\
	    \lstick{$\ket{0}$} & \qw & \gate[wires=3]{\mathcal{A}_0} & \targ{} & & & \qw \\
	    \lstick{$\ket{0}$} & \gate{X} & \qw & \gate[wires=2]{\mathcal{A}_2} & \qw & \qw & \qw\\
	    \lstick{$\ket{0}$} & \gate{X} & \qw & & \qw & \qw & \qw
	\end{quantikz}
	\vspace{-0.1in}
	\caption{Selection of potential placement locations.}
	\label{fig:placement}
	    \vspace{-0.2in}
\end{figure}

Algorithm~\ref{alg:mining} shows the assertion mining process. As shown in the algorithm, the first step is static analysis of the quantum circuit ($QC$) to identify placeholders ($PH$) for assertion mining. The static analysis will uncover the common functionalities such as inclusion of Hadamard gates ($H$), CNOT gates, X gates, etc. By uncovering the common functionalities, it is easy to identify which qubits are important to check in the circuit. These identified qubits are then randomly sampled from the quantum circuit. In other words, out of all the qubits and different placeholders, some are selected randomly. An example of random qubit and placeholder selection is shown in Figure~\ref{fig:placement}. For the first placeholder ($A_0$), only 3 qubits are selected and the measurement is performed immediately after the two X gates. The second placement ($A_1$) consists of one qubit and the measurement is conducted after the H gate. 

To perform the identification, we take inspiration from quantum circuit cutting \cite{peng2020simulating}. We formulate the act of measurement as a projective operation $O_i$, and so the probability value of measuring a set of qubits is $\mathbb{E}(\rho) = \mathrm {Tr}(O_i \rho)$ -- where $\rho$ is the density matrix of qubits. Consider an example consisting of single qubit, we may express the state of the qubit as 

\vspace{-0.1in}
$$
\rho = \frac{1}{2}\sum_{i=1}^{8} c_i Tr(O_i \rho)p_i
\vspace{-0.1in}
$$
where the Pauli matrices correspond to all the projective operators $O_i$, and $p_i$ are operators' eigenprojectors and their corresponding eigenvalues $c_i$. Following quantum circuit cutting, we may write the term as two quantum circuits: one for computing the expectation value $\mathrm{Tr}(\rho O_i)$, and the other for preparing the eigenstate $p_i$. By finding such occurrences in a quantum circuit, we can readily measure the expectation value, and once that is finished, we can simply reinitialize the state $p_i$ to continue the quantum computation. Due to this property, this is a good place to insert a quantum assert. To find these points practically, we utilize randomized circuit cutting by randomly inserting measure-and-prepare channels~\cite{lowe2022fast}.

As shown in line 5 in the algorithm, for each placeholder identified during static analysis, design instrumentation is conducted. During design instrumentation, we add measurement points for the selected qubits in the placement of placeholder. The sub quantum circuit ($SQC$) is used in the next steps of the assertion mining process. The next step is to simulate the $SQC$ using different inputs. For each iteration a random input is generated and the identified $PH$ is simulated using the random input. When simulating the $PH$ section ($SQC$) of the original circuit, it collapses the quantum state to classical state in the measurement points and the probability distribution of the outputs is stored in the execution trace. 

\begin{algorithm}[htp]
\caption{Assertion Mining}\label{alg:mining}
\begin{flushleft}
\hspace*{\algorithmicindent} \textbf{Input}  Quantum Circuit ($QC$), Iterations ($I$)\\
\hspace*{\algorithmicindent} \textbf{Output} Assertions ($A$)
\end{flushleft}
\begin{algorithmic}[1]
\State $PH$ $\gets $StaticAnalysis($QC$)
\State $A \gets \emptyset$
\For{$p \in PH$}
\State $trace \gets \emptyset$
\State $SQC \gets $Instrumentation($PH$)
\For{$i \in I$}
\State $input \gets $RandomInput()
\State $trace \gets trace \cup$ RandomSim($input$, $SQC$)
\EndFor
\State $assertion \gets $StatisticalAnalysis($trace$)
\State $A \gets A \cup assertion$
\EndFor
\State Return $A$
\end{algorithmic}
\end{algorithm}
\vspace{-0.1in}


After the simulation is completed for each placeholder and the iterations, the execution trace is analyzed to mine the assertions. We are using chi-squared testing to analyze the probability distributions in the execution trace. Chi-squared test is a statistical hypothesis test that is used to identify significant differences between expected frequencies with observed frequencies~\cite{mchugh2013chi}. We use chi-squared testing to determine whether an output probability distribution lies on the three classes of states: classical, uniform superposition, and bell state~\cite{huangStatisticalAssertionsValidating2019}. The results are used to generate the three types of assertions as described below.

\vspace{0.05in}
\noindent \textbf{Classical Quantum Assertions:} For classical assertions, the hypothesis of the chi-square test is selected such that the expected distribution should be a unimodal distribution with one peak value. 
After comparing the expected distribution with observed distribution, if the $p-value$ of the test is less than 0.05, the null hypothesis is rejected. This means that the observed distribution is not unimodal. If the $p-value$ is higher and closer to 1, the null hypothesis is accepted.

\vspace{0.05in}
\noindent \textbf{Uniform Superposition Quantum Assertions:} For these assertions, the hypothesis of the chi-square test is selected such that the expected distribution should be a uniform distribution. The uniform probability value is calculated using $1/2^n$, where n is the number of qubits. 
Since there are 2 qubits, the output can have 4 ($2^2$) states ($00$, $01$, $10$, $11$) and the probability of each state is roughly around 0.25 ($1/4$). If $p-value$ is less than 0.05, the null hypothesis is rejected, meaning the observed distribution is not uniform. Otherwise, if $p-value$ is greater than 0.05, the observed distribution is identified as uniform distribution. 

\begin{table}[htp]
\centering
\vspace{-0.05in}
\caption{Contingency table for Bell state} 
\label{tab:contingency}
\begin{tabular}{|cc|cc|}
\hline
\multicolumn{2}{|c|}{\multirow{2}{*}{Probability}}     & \multicolumn{2}{c|}{Second qubit} \\ \cline{3-4} 
\multicolumn{2}{|c|}{}                                 & \multicolumn{1}{c|}{0}     & 1    \\ \hline
\multicolumn{1}{|c|}{\multirow{2}{*}{First qubit}} & 0 & \multicolumn{1}{c|}{0.5}   & 0    \\ \cline{2-4} 
\multicolumn{1}{|c|}{}                             & 1 & \multicolumn{1}{c|}{0}     & 0.5  \\ \hline
\end{tabular}

\end{table}
\noindent \textbf{Entanglement Quantum Assertions:} For entangled assertions, chi-square test is performed on a contingency table. The expected distribution is represented in a contingency table as shown in Table~\ref{tab:contingency} where the probability is 0.5 when the first qubit is '0' followed by second qubit being '0' or first qubit is '1' followed by second qubit being '1'.  All the probabilities for other occurrences are 0. The chi-square test is performed between the observed distribution and the probabilities of contingency table to determine the Bell state.
If the $p-value$ is less than 0.05, the hypothesis is rejected, otherwise the quantum state is accepted to be a Bell state. 
 




\subsection{Assertion Implementation}
\label{sec:insertion}



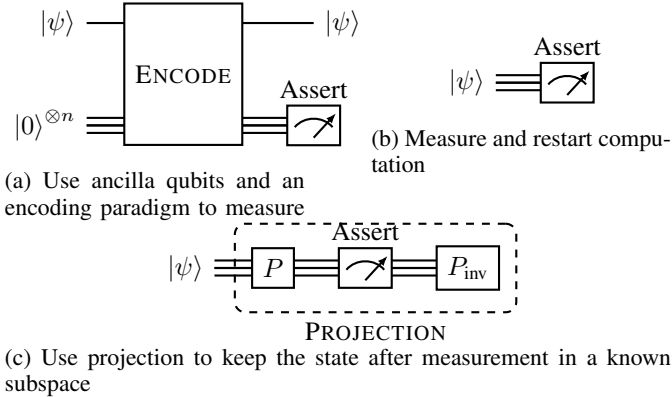
\begin{figure}[htp]
\vspace{-0.1in}
\centering
\begin{subfigure}{0.45\linewidth}
\centering
\begin{quantikz}
\lstick{$\ket{\psi}$} & \gate[wires=2]{\text{\sc Encode}} & \qw \rstick{$\ket{\psi}$}\\
\lstick{$\ket{0}^{\otimes n}$} & \qwbundle[alternate]{} & \meter{Assert}\qwbundle[alternate]{}
\end{quantikz}
\caption{Use ancilla qubits and an encoding paradigm to measure}
\label{fig:encode}
\end{subfigure}
\hfill
\begin{subfigure}{0.45\linewidth}
\centering
\begin{quantikz}
\lstick{$\ket{\psi}$} & \meter{Assert}\qwbundle[alternate]{}
\end{quantikz}
\caption{Measure and restart computation}
\label{fig:meas}
\end{subfigure}
\hfill
\begin{subfigure}{1\linewidth}
\centering
\begin{quantikz}
\lstick{$\ket{\psi}$} & \gate{P}\gategroup[1,steps=3,style={dashed,
rounded corners, inner xsep=3pt, inner ysep=5pt},background, label style={label position=below,anchor=north,yshift=-0.2cm}]{{\sc Projection}}\qwbundle[alternate]{} & \meter{Assert}\qwbundle[alternate]{} & \gate{P_{\text{inv}}}\qwbundle[alternate]{}
\end{quantikz}
\vspace{-0.1in}
\caption{Use projection to keep the state after measurement in a known subspace}
\label{fig:proj}
\end{subfigure}
\caption{Three alternatives for assertion implementations.}
\end{figure}

After assertion mining, we perform random sampling on assertions and implement the selected assertions. Implementation of assertions is an active research problem. Evaluating a property of a quantum state requires measurement, which will collapse the state to a basis state with some probability. Therefore, after conducting a measurement, it may not be possible to accurately continue the remaining quantum computation. To overcome this dilemma, assertions can be implemented using one of the following alternatives: 

\begin{enumerate}
    \item encoding the quantum state such that the measurement does not collapse the encoded state, as shown in Figure~\ref{fig:encode}.
    \item conducting a direct measurement and simply ignoring the rest of the computation, as shown in Figure~\ref{fig:meas}.
    \item using a projection-measurement that identifies if the quantum state is in a specified subspace of the total state space, as shown in Figure~\ref{fig:proj}.
\end{enumerate}

\begin{figure*}[htp]
  \centering
  \includegraphics[width=0.9\linewidth]{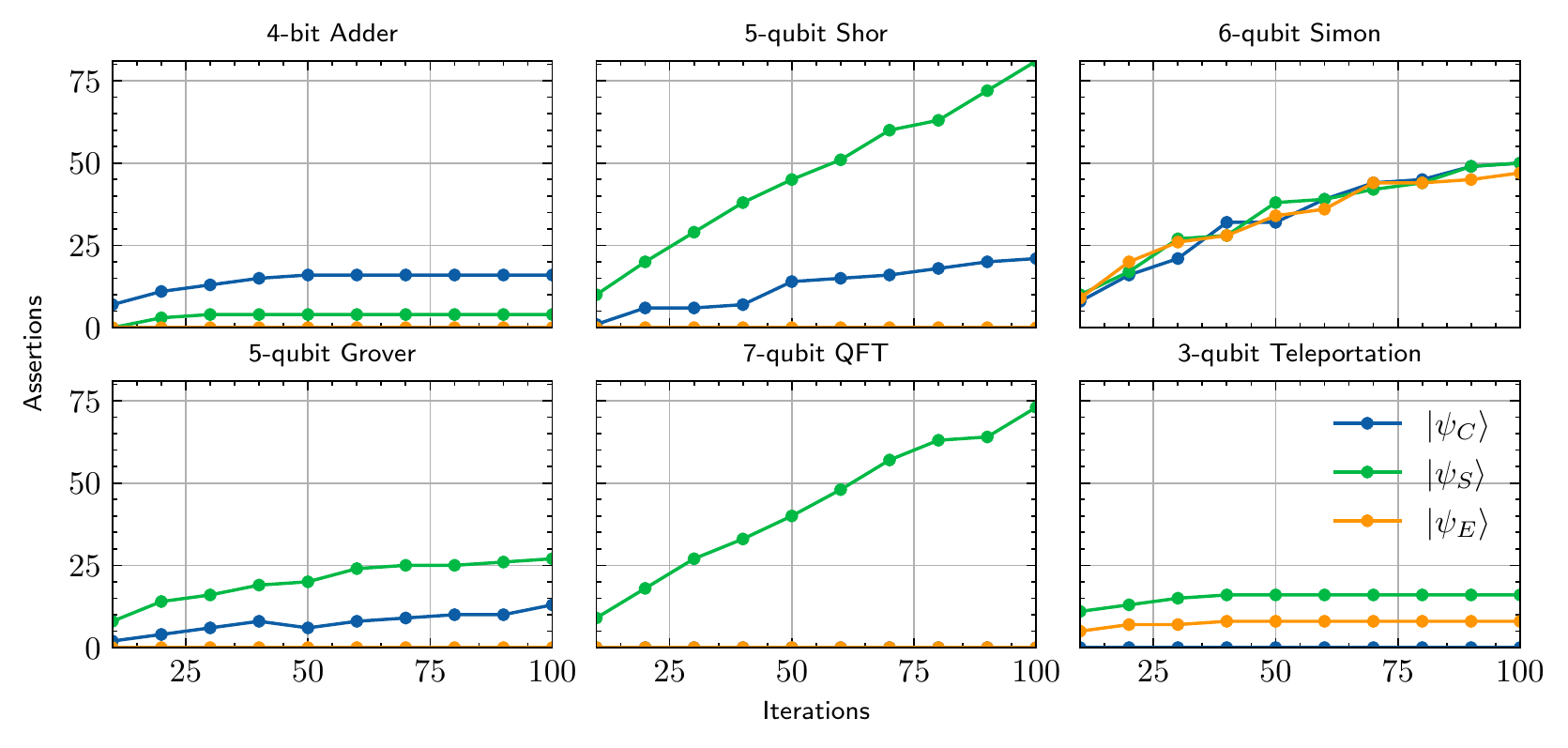}
  \vspace{-0.1in}
  \caption{Number of mined assertions and the corresponding number of simulations (iterations: $I$).}
  \vspace{-0.2in}
  \label{fig:results}
\end{figure*}

Each of the choices comes with its own disadvantages or overhead: (a) requires ancilla qubits and additional gates, (b) requires restarting the execution and removing the assertion to proceed with the remaining computation, and (c) requires classically constructing $2^n \times 2^n$ matrices which becomes infeasible for large subspaces.

However, we can optimize our assertion implementation by appropriately choosing the implementation that best suits the given assertion. For example, asserting a classical state may be best implemented by directly measuring, since the measurement of a classical state will simply collapse it to itself, and therefore not obstructing the remaining computation. Asserting an entangled state is well suited for projection-based implementation since the entangled state belongs to the stabilizer subspace which is determined by the stabiliser subgroup. Since the stabiliser subgroup is generated by at most $n$ Pauli operators, the implementation scales linearly.




%% file: section/experiment.tex
\section{Experiments}
\label{sec:experiments}

This section demonstrates the effectiveness of our automated assertion generation framework. First, we describe our experimental setup. Next, we present the results of assertion generation. Finally, we evaluate the quality of the generated assertions for error detection.  

\subsection{Experimental Setup}

For  experimental evaluation, we have selected six quantum circuits that are widely used in the quantum assertions community: 4-qubit adder, 5-qubit Shor's algorithm, 6-qubit Simon's algorithm, 5-qubit Grover's algorithm, 7-qubit QFT, and 3-qubit quantum teleportation algorithm. Our approach will work irrespective of the number of qubits in a design. The  assertion generation framework is implemented using Python and Qiskit. The classical simulation is performed using Aer simulator, utilizing low-rank stabilizer simulation for larger designs \cite{bravyiSimulationQuantumCircuits2019}. We ran our experiments on Intel i7-5500U @ 3.0GHz CPU with 16GB RAM machine.

\subsection{Experimental Results}

In this section, we present our experimental results for the six quantum circuits in three avenues: assertions mining, error coverage, and quality of assertions.

\subsubsection{Assertion Mining Results}

The design is first statically analyzed and randomly sampled to identify which qubits to measure and where to measure in the design. Then test vectors are generated randomly and the randomly sampled placeholder ($PH$) are simulated to get the execution traces. These traces are analyzed using statistical chi-square testing to automatically identify the quantum state at the measured instance. The sample size is $2^{13}$. Figure~\ref{fig:results} shows the number of assertions generated for the six quantum circuits while increasing the number of test inputs (simulation iterations). Assertion generation for all the three types of assertions, classical ($\ket{\psi_C}$), uniform superposition ($\ket{\psi_{S}}$) and cat entanglement ($\ket{\psi_{E}}$)), are shown in blue, green and orange colors, respectively. 

For example in case of the 4-qubit adder, total 16 classical assertions are mined using classical inputs with one placeholder after the whole process. Since the design has  4 inputs, the all possible classical assertions are 16 ($2^4$). The total 4 superposition assertions are mined using superposition inputs with one placeholder for the first two qubits after the whole process. The uniform superposition state can be observed only when the input of first two qubits is `11' (i.e., when input is 1, an H gate is applied to make the input superposition). Out of 16 possible inputs, only 4 of them have `11' for the first two qubits, which makes the 4 uniform superposition assertions. Similar process is conducted for the other circuits with different qubit selection and placeholders. 


\subsubsection{Coverage of Randomly Inserted Bugs}
\begin{figure}[htp]
	\centering
\includegraphics[width=0.8\columnwidth]{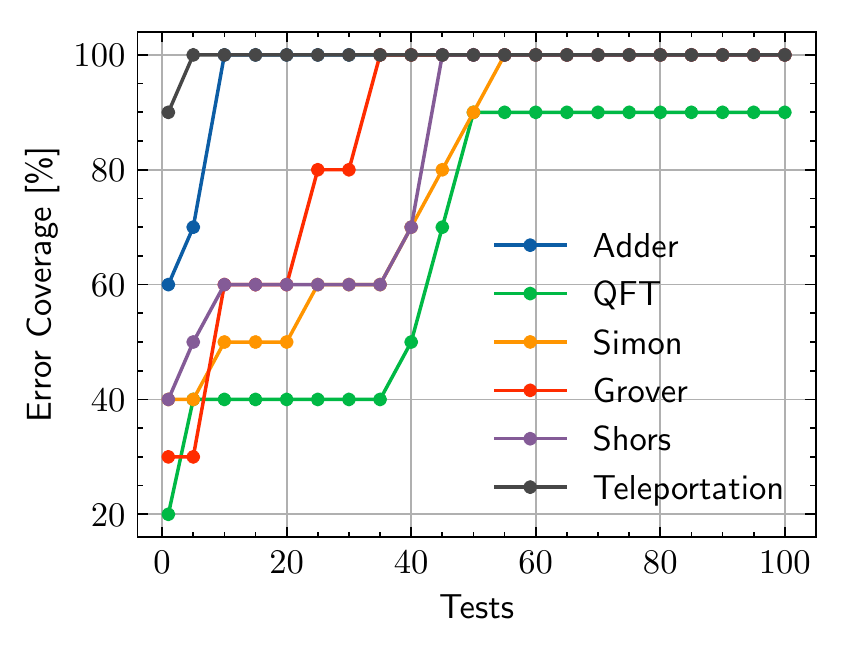}
 \vspace{-0.1in}
	\caption{Error coverage using increasing number of test patterns}
	\label{fig:error}
	    \vspace{-0.1in}
\end{figure}

It is important to verify the quality of the assertions that are mined from our framework. For this experiment, we randomly selected 5 assertions that got mined for each circuit and implemented the assertions as described in Section~\ref{sec:insertion}. Then 10 bugs were inserted in each design. These bugs include: change of existing gates to different gates, inclusion of new gates, removing some of the existing gates, doing random rotations to different qubits, etc. Figure~\ref{fig:error} presents the error coverage results with increasing test vectors for all the 6 quantum circuits. We generated random test vectors,  simulated them, and observed the error coverage for different test vectors. If any of the 5 assertions failed for the given test vectors due to an inserted error, then it was considered as an error coverage for that number of test vectors. As shown in Figure~\ref{fig:error}, small designs like adder (4-bits) and teleportation (3-bits), 100\% error coverage is achieved with less than 20 test cases. This is because any of the 5 assertions are likely to get activated since there are only 16 and 8 possible inputs for adder and teleportation circuit, respectively. For larger designs, like QFT with $2^7$ possible inputs and Simon's with $2^6$ possible inputs, higher error coverage is achieved with more test vectors. In the case of QFT, only 90\% of the errors are covered due to the fact that one error included in QFT did not change the uniform superposition state of the output.

\subsubsection{Trade-off Between Number of Assertions and Time}
\begin{figure}[htp]
	\centering
\includegraphics[width=0.8\columnwidth]{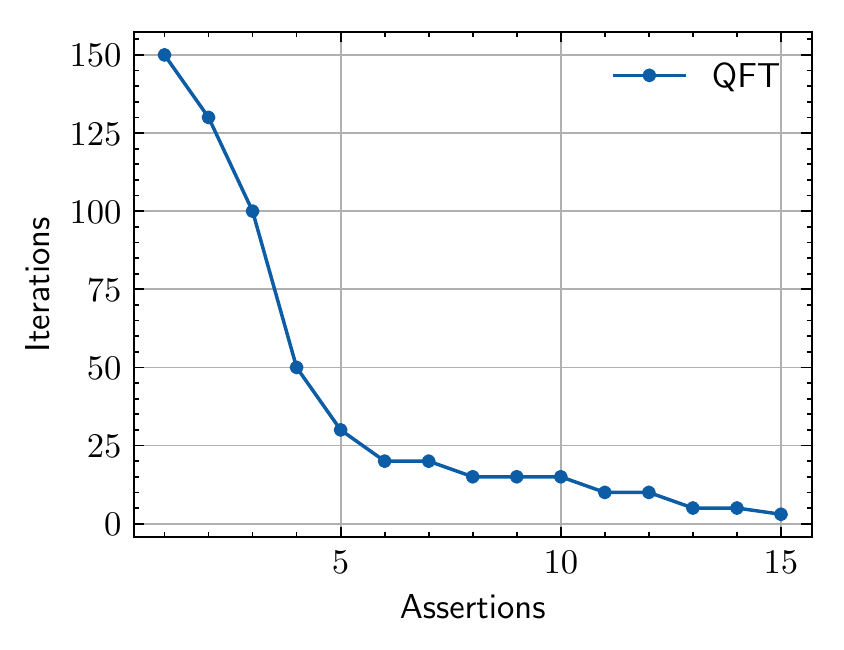}
 \vspace{-0.1in}
	\caption{Minimum iterations needed to identify a bug in 7-bit QFT while increasing the number of inserted assertions.}
	\label{fig:tradeoff}
	    \vspace{-0.2in}
\end{figure}

Insertion of assertions will increase the area overhead of a circuit due to the addition of new qubits and gates as described in Section~\ref{sec:insertion}. However if the number of inserted assertions is less, the error detection will take more simulation iterations with random inputs. This is shown in Figure~\ref{fig:tradeoff}. For this experiment, we have added one bug to the QFT design and kept increasing the number of inserted assertions. Then we check the number of iterations needed to identify the bug for varying number of inserted assertions. For example, when there is only one assertion inserted in the design, it took 150 random test vectors to identify the bug. This is because there are 128 different possible test vectors and only one of the test vectors match with the input of the assertion. For all the other test vectors, although the design is buggy, the assertion did  not get activated. With 15 inserted assertions, it took less than 10 test vectors to detect the bug. An important consideration is to find a trade-off between the number of assertions to insert in the circuit and the number of test vectors needed to activate the assertions. Depending on the design and time constraints, designers can choose the most profitable number of assertions to insert to obtain higher error coverage. 

%% file: section/conclusion.tex
\section{Conclusion}
\label{sec:conclusion}

Debugging a quantum circuit is a major challenge since it is harder to observe the internal states of quantum systems compared to its classical counterpart. One way to overcome this challenge is to use assertions to debug quantum circuits. While there are existing efforts in utilizing quantum assertions, they rely on expert knowledge of the designer to manually craft and place these assertions. In this paper, we presented a framework to automatically generate quantum assertions that has the capability to check three different quantum states: classical, superposition, and entanglement. Extensive experimental results demonstrated the effectiveness of our framework in generating high-quality assertions to detect bugs in diverse quantum circuits.